\input harvmac
\noblackbox
\def\Title#1#2{\rightline{#1}\ifx\answ\bigans\nopagenumbers\pageno0\vskip1in
\else\pageno1\vskip.8in\fi \centerline{\titlefont #2}\vskip .5in}

\def\msurr{\mathsurround=0pt}
\def\overleftrightarrow#1{\vbox{\msurr\ialign{##\crcr
	$\leftrightarrow$\crcr\noalign{\kern-1pt\nointerlineskip}
	$\hfil\displaystyle{#1}\hfil$\crcr}}}
\def\lrd{{\overleftrightarrow \partial}}

\def\gbdel{{G_{B\partial}}}

\def\vecx{{\vec x}}
\def\veck{{\vec k}}

\def\limrho{\buildrel \rho\rightarrow \pi/2 \over \longrightarrow}

\def\eg{{\it e.g.}}
\def\zhat{{\hat z}}

\def\lrd{\overleftrightarrow{\partial}}

\def\calO{{\cal O}}
\def\calA{{\cal A}}

\def\Ltilde{{\tilde L}}

\def\ehat{{\hat e}}
\def\Nhat{{\hat N}}

\def\tautilde{{\tilde \tau}}
\def\tauhat{{\hat \tau}}
\def\gtilde{{g}}
\def\Cprop{{C}}
\def\thetab{{\bar\theta}}
%
%
\def\ajou#1&#2(#3){\ \sl#1\bf#2\rm(19#3)}
\def\jou#1&#2(#3){,\ \sl#1\bf#2\rm(19#3)}
%
%

\font\ticp=cmcsc10
%
%

\lref\Juan{
  J.~M.~Maldacena,
  ``The large N limit of superconformal field theories and supergravity,''
  Adv.\ Theor.\ Math.\ Phys.\  {\bf 2}, 231 (1998)
  [Int.\ J.\ Theor.\ Phys.\  {\bf 38}, 1113 (1999)]
  [arXiv:hep-th/9711200].
}

\lref\GMH{
  S.~B.~Giddings, D.~Marolf and J.~B.~Hartle,
  ``Observables in effective gravity,''
  Phys.\ Rev.\  D {\bf 74}, 064018 (2006)
  [arXiv:hep-th/0512200].
}

\lref\Witt{
  E.~Witten,
  ``Anti-de Sitter space and holography,''
  Adv.\ Theor.\ Math.\ Phys.\  {\bf 2}, 253 (1998)
  [arXiv:hep-th/9802150].
}

\lref\GubserBC{
  S.~S.~Gubser, I.~R.~Klebanov and A.~M.~Polyakov,
  ``Gauge theory correlators from non-critical string theory,''
  Phys.\ Lett.\  B {\bf 428}, 105 (1998)
  [arXiv:hep-th/9802109].
}

\lref\HamiltonJU{
  A.~Hamilton, D.~N.~Kabat, G.~Lifschytz and D.~A.~Lowe,
  ``Local bulk operators in AdS/CFT: A boundary view of horizons and
  locality,''
  Phys.\ Rev.\  D {\bf 73}, 086003 (2006)
  [arXiv:hep-th/0506118].
}
\lref\BuLu{
  C.~P.~Burgess and C.~A.~Lutken,
  ``Propagators And Effective Potentials In Anti-De Sitter Space,''
  Phys.\ Lett.\  B {\bf 153}, 137 (1985).
}

\lref\Pol{
  J.~Polchinski,
  ``S-matrices from AdS spacetime,''
  arXiv:hep-th/9901076.
}

\lref\Sus{
  L.~Susskind,
  ``Holography in the flat space limit,''
  arXiv:hep-th/9901079.
}

\lref\bsm{
  S.~B.~Giddings,
  ``The boundary S-matrix and the AdS to CFT dictionary,''
  Phys.\ Rev.\ Lett.\  {\bf 83}, 2707 (1999)
  [arXiv:hep-th/9903048].
}

\lref\AbSt{
  M. Abramowitz and I.A. Stegun,
  {\sl Handbook of Mathematical Functions}, 
  Dover (NY) 1965.
}

\lref\Henrici{
  P. Henrici, 
  ``On Generating Functions of the Jacobi Polynomials'',
  Pacific J. Math. {\bf 5}, 923 (1955).
}

\lref\Dohse{
  M.~Dohse,
  ``Configuration Space Methods and Time Ordering for Scalar Propagators in (Anti and) de Sitter Spacetimes'',
  arXiv:0706.1887 [hep-th].
}

\lref\GGP{
  M.~Gary, S.~B.~Giddings and J.~Penedones,
  ``Local bulk S-matrix elements and CFT singularities,''
  arXiv:0903.4437 [hep-th].
}

\lref\GaGi{
  M.~Gary and S.~B.~Giddings,
  ``Relational observables in 2d quantum gravity,''
  Phys.\ Rev.\  D {\bf 75}, 104007 (2007)
  [arXiv:hep-th/0612191].
}

\lref\GiMa{
  S.~B.~Giddings and D.~Marolf,
  ``A global picture of quantum de Sitter space,''
  Phys.\ Rev.\  D {\bf 76}, 064023 (2007)
  [arXiv:0705.1178 [hep-th]].
}
\lref\GiMai{ S.~B.~Giddings and D.~Marolf, work in progress.}

\lref\GiSr{
  S.~B.~Giddings and M.~Srednicki,
  ``High-energy gravitational scattering and black hole resonances,''
  Phys.\ Rev.\  D {\bf 77}, 085025 (2008)
  [arXiv:0711.5012 [hep-th]].
}

\lref\GiPo{
  S.~B.~Giddings and R.~Porto, in preparation.
}

\lref\ReSi{
  M.~Reed and B.~Simon,
  ``Methods Of Mathematical Physics. Vol. 3: Scattering Theory,''
  {\it  New York, USA: Academic (1979) 463p}.
}

\lref\FSS{
  S.~B.~Giddings,
  ``Flat-space scattering and bulk locality in the AdS/CFT  correspondence,''
  Phys.\ Rev.\  D {\bf 61}, 106008 (2000)
  [arXiv:hep-th/9907129].
}

\lref\DulB{
  C.~Dullemond and E.~van Beveren,
  ``Propagators In Anti-De Sitter Space-Time,''
  J.\ Math.\ Phys.\  {\bf 26}, 2050 (1985).
}

\Title{
\hbox{
CERN-PH-TH/2009-051
}
}
{\vbox{\centerline {The flat space S-matrix from}
\vskip2pt\centerline{the AdS/CFT correspondence?}
}}
\centerline{{\ticp Michael Gary}\footnote{$^\dagger$}{Email address: mgary@physics.ucsb.edu}{\ticp \ and Steven B. Giddings}\footnote{$^\ddagger$}{Email address: giddings@physics.ucsb.edu}}
\vskip.05in
\centerline{\sl Department of Physics, University of California}
\centerline{\sl Santa Barbara, CA 93106-9530}
\centerline{\sl and}
\centerline{\sl PH-TH, CERN}
\centerline{\sl Geneva, Switzerland}

\bigskip
\centerline{\bf Abstract}
We investigate recovery of the bulk S-matrix from the AdS/CFT correspondence, at large radius.  
It was recently argued that some of the elements of the S-matrix might be read from CFT correlators, given a particular singularity structure of the latter, but leaving the question of more general S-matrix elements.  
Since in AdS/CFT, data must be specified on the boundary, we find certain limitations on the corresponding bulk wavepackets and on their localization properties.  In particular, those we have found that approximately localize have low-energy tails, and corresponding power-law tails in position space.  When their scattering is compared to
that of ``sharper" wavepackets typically used in scattering theory, one finds apparently significant differences, suggesting a possible lack of resolution via these wavepackets.  We also give arguments that construction of the sharper wavepackets may require non-perturbative control of the boundary theory, and particular of the $N^2$ matrix degrees of freedom.  These observations thus raise interesting questions about what principle would guarantee the appropriate control, and about how 
a boundary CFT can accurately approximate the flat space S-matrix.

\Date{}

\newsec{Introduction}

The proposed\refs{\Juan} AdS/CFT correspondence has inspired an enormous amount of investigation, particularly due to its claim to provide an equivalence between a boundary conformal field theory, and a non-trivial higher-dimensional bulk string theory.  As such, it has been deemed to hold the prospect of serving as a non-perturbative definition of string theory in AdS space.  Such a definition would in turn allow one to investigate all the mysteries of gravitational physics in such a space, and in particular questions such as how to give a unitary description of black hole formation and evaporation, {\it etc.}  Of course, the asymptotics of AdS are very different from Minkowski space, but the radius $R$ of AdS is a free parameter, and as such we could imagine adjusting it to be large as compared to all relevant physical scales, and thus imagine recovering, in a good approximation, string theory in flat space.  For that reason, there is widespread belief that the correspondence furnishes a complete description of physics in this limit.

However, in practice it has been much easier to read the correspondence in the direction of bulk to boundary, namely to infer from some given bulk phenomenon, an image phenomenon on the boundary.  It is comparatively more difficult to infer from the boundary theory the expected detailed structure of the bulk theory, which has been widely anticipated to behave approximately like a local field theory at distances large as compared to the string or Planck scales, and small as compared to the AdS radius.

A particularly important target is the bulk S-matrix.  The CFT correlators define an AdS analog of the S-matrix\bsm.  But, to realize the goal of using the CFT as a non-perturbative definition of string theory, and in particular to directly investigate various non-trivial features of quantum gravity, one would like to derive from the boundary theory a unitary bulk S-matrix describing scattering in the flat-space limit.  Early investigations of this question include \refs{\Pol,\Sus,\FSS}.

 Further progress was made on this matter  in a recent paper\refs{\GGP}.  In particular, it was found that for CFTs with a particular singularity structure in their correlators, there is a prescription to extract features of a bulk S-matrix, in the plane-wave limit.  One might be tempted to view this as answering  in the affirmative the question of whether the S-matrix can in principle be extracted.  However, one should recall that a careful definition of the S-matrix provides its matrix elements for a complete set of asymptotic states, which behave like asymptotically non-interacting multi-particle states in the Hilbert space of a free field theory.\foot{Here we are being slightly incautious about all the aspects of the LSZ formalism.}  The plane wave states are not states in a Hilbert space, although one can ordinarily superpose them to provide states of the Hilbert space.  But, if we consider the case of $R$ large as compared to all relevant physical scales, but finite, we do not exactly produce the plane wave states.  Specifically, we found in \GGP\ that we could extract plane-wave S-matrix elements from the $R\rightarrow\infty$ limit of wavepackets termed 
``boundary-compact.''  At finite $R$, we can define scattering via the boundary theory for the space of states generated by the boundary-compact states; more general states typically produce pathologies such as divergent behavior near the boundary of AdS\FSS.  

This raises the question of whether scattering amplitudes for boundary compact wavepackets contain all the information of the bulk S-matrix.  In exploring this, we will exhibit some differences between scattering amplitudes of boundary-compact wavepackets and those of  wavepackets usually used in  scattering theory.  These suggest that boundary-compact wavepackets could lack a certain degree of resolution of bulk physics.  However, this does not necessarily imply that boundary-compact wavepackets lack some of the scattering information, if one can reproduce healthier wavepackets from sums of boundary compact wavepackets.  While we do not have a complete answer whether this is possible, we investigate this question, and find arguments suggesting that recovery of more local wavepackets such as those of rapid falloff (Schwartz functions) or compact support may not automatically follow, in the absence of further non-perturbative input from the matrix degrees of freedom of the boundary theory.  This then raises the question of how such a boundary theory might adequately reproduce a unitary bulk S-matrix.

In outline, the next section will first sketch, as a review and to set the stage, the usual treatment of the S-matrix for scattering of wavepackets.  We then introduce the boundary-compact wavepackets and explore some of their properties in section three; in particular, we will show that they have long-range power-law tails in position space, which are closely associated with the corresponding boundary correlators.  Section four then considers scattering of these wavepackets, and shows that for certain of these wavepackets, the tails can make non-negligible contributions to the scattering amplitudes, such that they do not closely approximate those of wavepackets more commonly used in scattering theory.  In section five, we discuss the apparent limitations of resolution that thus arise if one is only allowed to probe the bulk theory via boundary compact wavepackets.  We also briefly investigate the question of whether boundary-compact wavepackets might be superposed to form wavepackets of better resolving power.  Section six contains further discussion and conclusions.

\newsec{The problem of extracting the S-matrix}

\subsec{Posing the problem}

The question we will explore is the extent to which the flat space S-matrix can be extracted, perhaps in an approximation as the AdS radius becomes large, from the boundary theory in a proposed AdS/CFT dual.  For the purposes of asking this question, we would like to assume that we have a means of computing all relevant quantities (correlators) in the boundary theory.  However, since in practice such calculations are intractable, we can explore it via a more modest approach.  Namely we suppose, as is commonly done, we {\it begin} with a local bulk theory, and then use it to {\it define} the boundary theory.  Then, we ask whether from the corresponding correlators of the boundary theory, calculated via the underlying bulk theory\refs{\Witt,\GubserBC}, one can reproduce the flat space S-matrix of the bulk theory.  Certainly if we begin with such a bulk theory and cannot recover it in full, that suggests it would be 
even more problematical to recover the full bulk dynamics from an intrinsic definition of a boundary theory.

Since the bulk theory is gravitational, the best understood observables are the elements of the S-matrix.\foot{However, it has been proposed that one may define other {\it proto-local} observables that approximate local observables\refs{\GMH,\GaGi,\GiMa,\GiMai} in certain situations, as is needed to address questions of cosmology, {\it etc.}}
Here we assume that the gravitational S-matrix actually exists; one finds improvement of the usual infrared problems in dimensions $D\geq5$. (For further discussion see \refs{\GiSr,\GiPo}.)

There are two essential ingredients in the usual definition of the S-matrix.  The first is the identification of the Hilbert space of asymptotic states corresponding to ``freely propagating'' particles.  The second is the set of amplitudes for scattering from any given initial asymptotic state, to any given final asymptotic state.  
Given that for any finite AdS radius $R$ one is dealing with very different asymptotics from Minkowski space, it is important to take some care with the corresponding definitions in AdS.  Specifically, in many modern treatments of the S-matrix, one immediately takes the plane wave limit, corresponding to working in infinite volume.  However, such a cavalier procedure seems potentially problematic in the AdS context, since the geometry at large distance differs an arbitrarily large amount from that of flat space.

Instead, we will adhere to a more cautious approach to the S-matrix, using bona-fide states of the Hilbert space, {\it i.e.} normalizable wavepackets, as is done in careful treatments in flat space.  Specifically, if one works with wavepackets corresponding to normalizable states, the flat-space S-matrix is a rule that constructs finite amplitudes for scattering from a complete set of such incoming wavepacket states to a complete set of outgoing wavepacket states.
 
We would like to know whether the boundary amplitudes reproduce this quantity, at least to a very accurate degree of approximation as $R$ becomes large as compared to relevant scales.  Thus, this question has two parts.  The first part is whether, using quantities to which we have access if we work within the boundary theory, we can construct good approximations of the appropriate normalizable flat-space wavepackets.  The second part is whether the boundary theory yields amplitudes for their scattering, that well-approximate the amplitudes that follow from the flat space S-matrix.  Thus, in our view an important  question is {\it  whether  the boundary theory yields a  sufficiently complete set of normalizable states for which we get an accurate approximation to the flat space S-matrix.}

\subsec{Scattering in flat space}

Before investigating approaches to the S-matrix from AdS space, let us recall some basic features of the flat space story.  
In a perturbative expansion we typically write the S-matrix in the form
\eqn\Sdef{
S=1 +i{\cal T}\ .}
If we consider for example two-to-two particle scattering, in $D=d+1$ dimensions, the transition matrix $T$ then arises from the plane-wave matrix elements,
\eqn\Tdef{\langle p_3,p_4| {\cal T}|p_1,p_2\rangle = (2\pi)^D \delta^D(\sum_ip_i) { T}(p_i)\ .}
In the case of t-channel exchange of a massless particle with coupling $g$, we have a leading Born contribution
\eqn\TBorn{{T}(p_i) =- {g^2\over t}\ ;}
in the case of gravity, in the high-energy limit, $g^2 \sim G s^2$.

In a more careful approach where the S-matrix is defined on the Hilbert space of states, we recall that
eqs.~\Sdef, \Tdef\ define a distribution on that space.  
The scattering amplitudes arise from integrating the expressions \Sdef, \Tdef\ against the corresponding wavefunctions.  To be very explicit, if we have four wavepackets $\psi(p_{i0},\Delta p_i)$ with approximately definite momenta $p_{i0}$, and with momentum uncertainties $\Delta p_i$, then the scattering contribution to the amplitude is  of the form
\eqn\psiscatt{{\cal A}_s(p_{i0}, \Delta p_i)=i \int\prod_i \widetilde{dp}_i \psi(p_{i0},\Delta p_i;p_i) (2\pi)^D \delta^D(\sum_ip_i) {T}(p_i)\ ,}
where 
\eqn\LIvol{\widetilde{dp} = {d^{d}p\over (2\pi)^{d} 2 \omega_p}}
is the Lorentz-invariant volume element.  The scattering amplitude ${\cal A}_s(p_i, \Delta p_i)$ of course depends non-trivially on all of the $p_{i0}$ and $\Delta p_i$; there is also a direct contribution ${\cal A}_d$ to the full amplitude, that comes from the ``1'' in \Sdef.

There are different choices of wavepackets that are found in the literature.  One typically uses Schwartz functions, which fall more rapidly than any power of the distance from the center of the wavepacket.  A more refined choice is to chose wavepackets of compact support (which are of course Schwartz), either in position space or momentum space.\foot{The Fourier transform of a Schwartz function is Schwartz.}  The latter, wavefunctions of compact support in momentum space, and not including zero frequency, are used in {\it e.g.} \refs{\ReSi} and are referred to as {\it regular wavepackets}.

Such regular wavepackets in particular are useful for taking the plane wave limit, where the scattering amplitudes \psiscatt\ approach the form \Tdef.  To understand this limit better in terms of wavepackets, note that if in 
\psiscatt\   we consider regular wavepackets with very narrow support in momenta, $\Delta p_i \ll p_{i0}$, then this amplitude will vanish if $\sum_i p_{i0}$ are more than an amount $\sim \Delta p_i$ outside the momentum-conserving range.  If the $p_{i0}$'s approximately satisfy momentum conservation, then the scattered amplitude becomes
\eqn\Aapprox{{\cal A}_s(p_i, \Delta p_{i0}) \approx i{T}(p_{i0}) \int \widetilde{dp}_1 \widetilde{dp}_2 {p_3^{D-3}\over 4 E_4} {d\Omega_3\over (2\pi)^{D-2}} \psi_4\psi_3\psi_2\psi_1\ .}
%
%
%
%
where ${\vec p}_4 = {\vec p}_1+{\vec p}_2-{\vec p}_3$ and $E_3 = E_1 + E_2 - E_4({\vec p}_4)$, and $\Omega_3$ represents the angles of ${\vec p}_3$ with respect to some axis. 
Recall that this is not the full amplitude; there is also the direct contribution ${\cal A}_d$ from the ``one'' in \Sdef.  However, this does not contribute to the amplitude if we take $p_{30}$ and $p_{40}$ to be more than $\Delta p_i$ away from the ``forward'' direction,\foot{In cases with identical particles, one must also exclude the case with $p_{30}\leftrightarrow p_{40}$.} $p_{30}=p_{10}$, $p_{40}=p_{20}$.

\subsec{Approaches from AdS space}

In trying to extract the S-matrix from the boundary CFT, one approach is to seek directly an expression that gives a good approximation to the expressions \Sdef, \Tdef, \TBorn.  However, one finds\refs{\FSS} that there are important contributions from interactions near the boundary of AdS, which interfere with such a direct approach. 

This indicates the need to construct   wavepackets more carefully, as suggested in \refs{\Sus,\Pol}.  Indeed, since we want to extract the flat space S-matrix at large $R$,  we would like to consider a situation where our wavepackets interact only in one AdS region, or more specifically, following \Sus, in a small subregion of size $L\ll R$ of that AdS region, which we might refer to as the ``lab.''  

A first approach to this fails, as emphasized in \FSS.  Sources at the boundary correspond to non-normalizable bulk states.  If one considers interactions of generic non-normalizable bulk states, one finds that the integration over one of the interaction vertices of the t-channel Feynman diagram takes the form:
\eqn\NNint{\int dV \psi_{NN} \psi_{NN} G_B\ ,}
where $dV$ is the volume element in AdS, $\psi_{NN}$ denote the non-normalizable states, and $G_B$ is the bulk propagator.  This integral generically has a divergence near the boundary of AdS, far outside the lab region.

It seems clear that one wants to work with {\it normalizable} wavefunctions; indeed, these are the wavefunctions that will reduce to normalizable states in the flat space limit.  

There are two approaches to constructing normalizable wavefunctions in AdS.  The first is to specify normalizable initial data at timelike infinity.  One might hope to extract the flat space S-matrix from scattering of such states.  However, one immediately notes a significant problem.  AdS behaves like a finite-size box of size $R$, with reflecting walls.  If one introduces two wavepackets at $t=-\infty$, they will scatter an {\it infinite} number of times before they reach $t=+\infty$.  Thus, one apparently cannot isolate the contribution of a single scattering event, where the particles approach once from infinity, scatter, and subsequently separate.

In order to construct wavepackets that only scatter once, one would thus like to send in states from the spacelike boundary of AdS.  This returns one to the problem of introducing sources on the boundary, and consequent non-normalizable behavior.  However, one might attempt to limit the damage, by taking the boundary source to be of compact support.  In this case, the corresponding bulk wavefunction will be normalizable for AdS times outside the range of times of the support.  We will refer to such states, which were recently introduced in \GGP,  as {\it boundary-compact} states.\foot{Compact sources have also been studied in related contexts in \refs{\Witt,\HamiltonJU}.}   Using this type of state is the only obvious solution to the problem of constructing bulk normalizable wavepackets that localize and scatter only once.

We next turn to a careful study of the construction of such wavepackets, and then to the question of their scattering amplitudes.

\newsec{Boundary-compact wavepackets and their properties}

\subsec{AdS preliminaries}

We begin with some basic aspects of AdS.  For more details, and notation, see {\it e.g.} \FSS.

We work in global coordinates for AdS${}_{d+1}$, $(\tau,\rho,\ehat)$, where $\ehat$ is a $d$-dimensional
unit vector.  The metric is
\eqn\glocoora{ds^2 = {R^2\over \cos^2\rho} ( -d\tau^2 + d\rho^2 +
\sin^2\rho \, d\Omega^2_{d-1} )\ ,}
and the AdS boundary corresponds to $\rho = \pi/2$.

We will also want to take the flat space limit of this metric; without loss of generality, we can work in the vicinity of the point $\tau=\rho=0$.  The flat space limit can then be explicitly seen in the coordinates 
\eqn\minkcoorda{t=R\tau\ ;\  r=R\rho\ ,}  
where the metric takes the form
\eqn\appflata{ds^2 = {1\over \cos^2(r/R)}\left[ -dt^2 + dr^2 + R^2
\sin^2\left({r\over R}\right)^2 d\Omega^2\right]\ .}
This is manifestly flat in the $R\rightarrow \infty$ limit.

\subsec{Construction of boundary-compact wavepackets}

As outlined in the preceding section, in order to construct wavepackets that are normalizable and thus have a good flat space limit, we use boundary sources that are compactly supported\GGP.  Recall some basic features of the AdS/CFT correspondence.  Let $\phi$ be a bulk field of mass $m$, and $\calO$ be its corresponding boundary operator.  Then, a boundary correlator including $\calO$ is related to a bulk correlator by taking the field to the boundary and rescaling:
\eqn\boundcorr{\langle\calO(b)\cdots\rangle =  2\nu R^{(d-1)/2}\lim_{\rho\rightarrow\pi/2} (\cos\rho)^{-2h_+}\langle \phi(\tau,\rho,\ehat)\cdots\rangle\ ,}
where $b=(\tau,\ehat)$ denotes the boundary coordinate and\foot{The quantity $2h_+$ is denoted $\Delta$ in \GGP\ and elsewhere in the AdS/CFT literature.}
\eqn\hpmdef{4h_{\pm} = d \pm\sqrt{d^2 + 4m^2R^2} = d\pm 2\nu\ .}

We can project on a particular wavepacket by integrating this expression against a boundary source $f(b)$.  In the Feynman diagrams used to compute the bulk correlator on the right hand side of \boundcorr, the Feynman propagator $G_B(x',x)=i\langle T\phi(x')\phi(x)\rangle$ thus gets replaced by 
\eqn\bdywavepack{\psi_f(x) =\int db' f(b') G_{B\partial}(b',x) }
where
\eqn\gbbdef{\gbdel(b',x) = 2\nu R^{(d-1)/2} \lim_{\rho'\rightarrow\pi/2}(\cos\rho')^{-2h_+}
G_B(x',x)\ }
is the bulk-boundary propagator.\foot{Note that the normalization here is missing a factor $R^{(d-1)/2}$ relative to the conventions of, \eg, \FSS.  The current convention is chosen to better exhibit the dimensionality of bulk vs. boundary objects.}  Thus, $\psi_f$ is the wavepacket resulting from the source $f$.  Note that one can then easily show that the wavepacket has non-normalizable behavior,
\eqn\psinn{\psi_f(x)\limrho {(\cos\rho)^{2h_-}\over R^{(d-1)/2}} f(b)\ .}

Boundary-compact sources are those with compact $f(b)$.  We will build such wavepackets by using a basic smooth 
function $L(x)$ of compact support, chosen so $L(0)=1$, and to vanish for $|x|>1$.  
We also would like to approximate as closely as possible incoming wavepackets of definite frequency.  Thus, we take wavepackets with additional time dependence:
\eqn\bdysource{f(b) =L\left({\tau-\tau_0\over \Delta \tau}\right) L\left({ \theta\over \Delta \theta}\right)  e^{-i\omega R (\tau-\tau_0)}\ ,}
where
$\cos\theta = {\hat e}\cdot {\hat e}_0$
and where $\Delta\tau$ and $\Delta \theta$ are characteristic widths.
Note from \minkcoorda\ that the frequency dependence introduced corresponds to frequency $\omega$ with respect to the time $t$.  These, together with \bdywavepack\ will define our basic boundary-compact wavepackets.  These wavepackets were used in \refs{\GGP} to extract S-matrix elements in the plane wave limit. Here we consider the more general problem of extracting a more complete specification of the S-matrix via such wavepackets.

We next turn to an investigation of the properties of these wavepackets.  A first question is how closely they approximate the desired regular or Schwartz wavepackets of flat space.  Also, when we investigate scattering, as we will see another relevant question is the appearance of the boundary ``image'' of these wavepackets, namely the quantity
\eqn\bdyopval{\langle 0| \calO(b) |\psi_f\rangle}
in the state created by the source $f(b)$; this quantity enters into the calculation of the direct contribution to scattering, analogous to $\calA_d$ in our discussion of flat space.  We will find that these questions are related.

\subsec{Wavepackets and  tails}

We are interested in properties of the boundary-compact wavepackets, of the form \bdywavepack, with sources as in \bdysource, particularly on distance scales short as compared to $R$.  Properties of the relevant Green functions are described in the appendix, and in particular, the bulk-boundary propagator is given in (A.7). For our purposes, it is convenient to rewrite this expression in a Schwinger-like form: 
\eqn\gbdelr{G_{B\partial}(b',x) = { (\cos\rho)^{2h_+}\Nhat \over R^{(d-1)/2} }\int_0^\infty d\alpha \alpha^{2h_+-1} \exp\left\{i\alpha [\cos(|\tau-\tau'|-i\epsilon) - \sin\rho\, \ehat\cdot\ehat']\right\}\ ,}
where 
\eqn\Nhatdef{\Nhat = {2\nu\Cprop\over i^{2h_+-1} \Gamma(2h_+)} }
is a normalization constant.  This simplifies the problem of inferring the behavior of the wavepacket.

For exploring distances $\ll R$, we need wavepackets sufficiently focussed in angle and time.  Thus, as in \GGP, we take $\Delta \tau$ and $\Delta \theta$ in \bdysource\ small.  In particular, in bulk coordinates, the longitudinal spread of the wavepacket is $\Delta t\sim \Delta \tau R$.  Thus, for a finite bulk wavepacket, $\Delta \tau$ should vanish in the large $R$ limit.

For such a narrowly peaked source $f(b')$, we can expand the exponent in \gbdelr\ about the central value $(\tau_0,\ehat_0)$, which we take to be $(-\pi/2,-\zhat)$ where $\hat z$ is the direction of propagation.
Thus, with $\tau'=\tautilde-\pi/2$ and $\tau>\tau'$, we have
\eqn\cosexp{\cos(|\tau-\tau'|-i\epsilon) \approx -\sin\tau + \cos\tau(\tautilde+i\epsilon) + \calO(\tautilde^2)\ .}
In the wavepacket \bdywavepack\ we then find
\eqn\tauint{\int d\tautilde L(\tautilde/\Delta \tau) e^{-i\omega R \tautilde} e^{i\alpha[\cos(|\tau-\tau'|-i\epsilon)] }\approx \Delta \tau e^{-i\alpha \sin\tau} \Ltilde[\Delta\tau(\omega R -\alpha \cos\tau )]\ ,}
where 
\eqn\Ltildedefa{\Ltilde(\kappa) = \int dx e^{-i\kappa x} L(x)}
is the one-dimensional Fourier transform of $L$.
This gives for the wavepacket \bdywavepack\
\eqn\wavepackb{\eqalign{\psi_f(x) \approx  { (\cos\rho)^{2h_+}\Delta\tau \over R^{(d-1)/2}} \Nhat\int_0^\infty d\alpha& \alpha^{2h_+-1} e^{-i\alpha \sin\tau} \Ltilde[\Delta\tau(\omega R -\alpha \cos\tau )] \cr &\int d^{d-1}e' e^{-i\alpha\sin\rho \ehat\cdot\ehat'}L({\theta'\over\Delta\theta})\ ,}}
where
\eqn\thetapdef{\cos\theta'=-\ehat'\cdot\zhat\ .}

Let us consider the behavior of this wavepacket in a flat region 
\eqn\flatlim{ |r|\ll R\quad , \quad |t|\ll R\ ;}
 here $\sin\rho\approx r/R$ and $\sin\tau\approx t/R$.  Define ${\vec r} = r \ehat$
%
and define the new variable
\eqn\kdef{k=\alpha/R\ .}
In the limit \flatlim, the wavepacket becomes
\eqn\flatpack{ \psi_f(x)\approx {\Delta t \Nhat R^{\nu-1/2} }  \int_0^\infty k^{d-1}dk d^{d-1}e' \left\{k^{2h_+-d}
\Ltilde[\Delta t(k-\omega)] L\left({\theta'\over\Delta\theta}\right)\right\} e^{-ik\ehat'\cdot {\vec r}-ikt} \ .}
This clearly exhibits the Fourier representation of the wavepacket, with momentum ${\vec k}=-k\ehat'$.

From this expression, one can immediately see several features of the bulk wavepackets arising from the boundary-compact sources.  First, note that since $L$ is compact support, its Fourier transform $\Ltilde$ is not, and therefore does not vanish at $k=0$. This means that the wavepackets are not regular.  Moreover, they are
not Schwartz. This follows from the fact that the wavepacket is not a smooth function of $\vec k$ at $k=0$, even with $2h_+=d$.  In fact, \flatpack\ explicitly exhibits a long-wavelength tail, at $k\ll \omega$, with size $\Ltilde(\omega\Delta t)$.  This tail is moreover $R$-independent.

One can also see these properties directly in position space.  
let us decompose 
\eqn\ehatpda{
\ehat' = -\cos\theta' \zhat + \sin\theta' \ehat'_\perp \approx -\zhat + \theta'\ehat'_\perp ,}
where $\ehat_\perp' \perp \zhat$, and where in the second equality we have used the fact that the source $f$ restricts one to small $\theta'$.  Next, note that for small $\theta'$, the integration measure is approximated 
\eqn\approxmeas{k^{d-1}d^{d-1}e' \approx k^{d-1} \theta^{\prime d-2}d\theta' d^{d-2}e'_\perp\approx d^{d-1}k_\perp\ }
where $k_\perp$ is interpreted as the transverse momentum.  This angular integral then gives
\eqn\angint{\int d^{d-1}k_\perp L(k_\perp/k\Delta\theta) e^{ik_\perp\cdot x_\perp} = (k\Delta\theta)^{d-1}
\Ltilde_{d-1}(k\Delta\theta x_\perp)\ ,}
the $d-1$-dimensional Fourier transform, written in terms of the transverse coordinate $x_\perp$.  Thus, the wavepacket is
\eqn\wavepackt{\psi_f(x)\approx {\Delta t (\Delta \theta)^{d-1}\Nhat R^{\nu-1/2} } \int_0^\infty dk k^{2h_+-1} \Ltilde[\Delta t(k-\omega)] \Ltilde_{d-1}(k\Delta\theta x_\perp)e^{ik(z-t)}\ .}
For small $x_\perp$ and $u=t-z$, the first $\Ltilde$ is the most sharply peaked term, and we find falloff of the wavefunction with characteristic behaviors 
\eqn\nearfall{\psi_f \approx \psi_f(0){\Ltilde_{d-1} (\omega \Delta\theta x_\perp)\over\Ltilde_{d-1}(0)} L(u/\Delta t)e^{-i\omega u}}
where
\eqn\psizero{\psi_f(0)={2\pi \Nhat(\Delta\theta)^{d-1}R^{\nu-1/2}}\omega^{2h_+-1}\Ltilde_{d-1}(0)\ }
is the value of the wavepacket at the origin.  (Notice that, if we want a wavepacket that is unit-norm in the bulk, we should rescale the source, \bdysource, and thus $\psi_f(0)$, by a coefficient $\propto R^{1/2 -\nu}$.)

When $x_\perp> \Delta t/\Delta \theta,\, u/\Delta \theta$, the second $\Ltilde$ is most sharply peaked.  This yields a {\it power law} tail,
\eqn\transtail{\psi_f \approx \psi_f(0){\omega \Delta t\Ltilde(\omega \Delta t){\hat L}\over(\omega\Delta\theta x_\perp)^{2h_+}}\ ,}
where $\hat L$ is a constant determined by the function $L$,
indicating that the function is indeed not Schwartz.  Likewise, the expression exhibits a tail at $u\gg \Delta t$:
\eqn\utail{ \psi_f \approx  {\psi_f(0)}{\omega \Delta t\Ltilde(\omega \Delta t)\over (\omega u)^{2 h_+}}{\Gamma(2h_+)\over 2\pi i^{2h_+}}\ .}

We can thus describe the characteristics of the wavepackets as follows.   In the longitudinal direction they have characteristic width $\sim \Delta t$, but power law falloff \utail, from the long-wavelength tail, outside that.  In the transverse direction, the wavepackets fall off exponentially in $x_\perp$ with width $\Delta x_\perp \sim 1/(\omega\Delta\theta)$, but for $x_\perp>\Delta t/\Delta \theta$, this becomes the power law falloff \transtail.  Note that $\Delta x_\perp \omega \sim {\cal O}(1)$ requires $\Delta \theta\sim {\cal O}(1)$.

\subsec{Boundary description of wavepackets}

The presence of the tails \transtail\ is in fact directly connected to the boundary description of the wavepacket.  To see this, note that in the latter description, the wavepacket should be visible through the boundary operator as in \bdyopval.  It is straightforward to derive this from \wavepackb, using the prescription \boundcorr\ for deriving the boundary correlators from those of the bulk.  By this means, one finds 
\eqn\boundopeval{\eqalign{\langle0|\calO(b)|\psi_f\rangle 
\approx {2\nu \Delta \tau \Nhat} \int_0^\infty d\alpha& \alpha^{2h_+-1} e^{-i\alpha \sin\tau}\cr&  \Ltilde[\Delta\tau(\omega R -\alpha \cos\tau )] \int d^{d-1}e' e^{-i\alpha \ehat\cdot\ehat'}L\left({\theta'\over \Delta \theta}\right)\ .}}
Now, the change of variables \approxmeas\ gives (compare \wavepackt)
\eqn\boundopb{\eqalign{\langle0|\calO(b)|\psi_f\rangle \approx 2\nu \Delta \tau (\Delta \theta)^{d-1} \Nhat \int_0^\infty d\alpha &\alpha^{2h_+-1} e^{i\alpha(\cos\theta -\sin\tau)} \cr&\Ltilde[\Delta\tau(\omega R -\alpha \cos\tau )] \Ltilde_{d-1}(\alpha\Delta\theta \sin\theta)\ .}}

For $\cos\tau/\sin\theta \gg \Delta\theta/\Delta\tau = R\Delta \theta/\Delta t$, the first $\Ltilde$ is most peaked in $\alpha$ and essentially enforces $\alpha\cos\tau \approx \omega R$.  
But, for $\cos\tau/\sin\theta \ll  R\Delta \theta/\Delta t$, the second $\Ltilde$ is most peaked, and enforces $\alpha\approx0$.  This range of $\theta$ corresponds to the region of the boundary $S^{d-1}$ away from the source point or its antipodal point.  In this region the operator behaves as
\eqn\optail{\langle0|\calO(b)|\psi_f\rangle \approx2\nu \Delta\tau(\Delta\theta)^{d-1}\Nhat {\Ltilde({\omega \Delta t})} \int_0^\infty d\alpha {\alpha}^{2h_+-1} e^{i\alpha(\cos\theta -\sin\tau)} \Ltilde_{d-1}(\alpha\Delta\theta \sin\theta)\ .}
This describes a ``signal'' of strength $\propto \Ltilde({\omega \Delta t})$ propagating along the trajectory $\cos\theta=\sin\tau$.  It has a characteristic width $(\cos\theta -\sin\tau)\roughly<\Delta\theta\sin\theta$.  For much larger values of $\cos\theta -\sin\tau$, its behavior can be found by rescaling $\alpha\rightarrow\alpha/(\cos\theta -\sin\tau)$.  This yields a power law falloff, 
%
\eqn\oppower{\langle0|\calO(b)|\psi_f\rangle\approx 4\pi \nu  \Delta \tau (\Delta\theta)^{d-1} \Nhat{\hat L} \Ltilde_{d-1}(0)
{\Ltilde({\omega \Delta t}) \over ( |\cos\theta -\sin\tau|)^{2h_+}}\ .} 
This  is very similar to \transtail\ --
the bulk tail is directly related to this propagating boundary signal along the $S^{d-1}$.  

Let us also work out the  behavior of the boundary operator near the antipode to the source, 
at $\theta=0,\tau\sim\pi/2$, where the direct contribution to the S-matrix, coming from the ``1'' in \Sdef, is important. We do this using the boundary propagator, given in appendix A, which follows from a second rescaling as in \gbbdef.  With $\tauhat=\tau-\pi$, we have
\eqn\cosantip{\cos(|\tau-\tau'|-i\epsilon) \approx -1 + {(\tauhat-\tau'-i\epsilon)^2\over2}\ .}
Define $\cos{\thetab} = -\ehat\cdot\ehat'$.  Then the boundary propagator near the antipodal point becomes
\eqn\antipprop{G_{\partial}(b,b')\approx{i\Cprop (2\nu)^2  2^{2h_+} \over [\tauhat-\tau'-2\sin(\thetab/2)-i\epsilon]^{2h_+}[\tauhat-\tau'+2\sin(\thetab/2)-i\epsilon]^{2h_+}}\ .}
Again rewriting the propagator in a Schwinger-like form, one finds
\eqn\antippropsh{G_{\partial}(b,b')\approx 
{{\hat N}^2 2^{2h_+}\over iC}
\int_0^\infty d\alpha_1d\alpha_2(\alpha_1\alpha_2)^{2h_+-1}
e^{-i\alpha_1[\tauhat-\tau'+2\sin{\thetab\over2}]-i\alpha_2[\tauhat-\tau'-2\sin{\thetab\over2}]}\ .}
Integrating the propagator \antippropsh\ against the source \bdysource, we will need
\eqn\tauintantip{\int d\tau'L(\tau'/\Delta\tau)e^{-i(\omega R-\alpha_1-\alpha_2)\tau'}=\Delta\tau\Ltilde\left[\Delta\tau(\omega R-\alpha_1-\alpha_2)\right]\ .}
We can also write $\thetab$ in terms of $\theta,\theta'$, and $\phi'$, where  $\cos\theta=\ehat\cdot\zhat$, $\cos\theta'=-\ehat'\cdot\zhat$, and where $S^{d-1}$ is described as an $S^{d-2}$ fibration over the interval, with $\phi'$  the angle of $\ehat'$ on the $S^{d-2}$ relative to $\ehat$. Expanding in $\theta'$, we find
\eqn\thetabexp{2\sin{\thetab\over2} \approx 2\sin{\theta\over2} -\theta' \cos{\theta\over2}\cos\phi'\ .}
This can then be used in the integral over angles to yield
\eqn\angleintantip{\int d^{d-1}\ehat'L(\theta'/\Delta\theta)e^{-2i(\alpha_1-\alpha_2)\sin{\thetab\over2}}\approx(\Delta\theta)^{d-1}\Ltilde_{d-1}\left[\Delta\theta(\alpha_1-\alpha_2)\cos{\theta\over2} \right]e^{-2i(\alpha_1-\alpha_2)\sin{\theta\over2}}}
where $\Ltilde_{d-1}$ is as in \angint. 
%
Combining these factors yields
\eqn\boundantip{\eqalign{\langle0|\calO(b)|\psi_f\rangle\approx {\Delta\tau(\Delta\theta)^{d-1}\Nhat^2 2^{2h_+}\over iC}&\int_0^\infty d\alpha_1d\alpha_2(\alpha_1\alpha_2)^{2h_+-1}\Ltilde\left[\Delta\tau(\omega R-\alpha_1-\alpha_2)\right]\cr&e^{-i\tauhat(\alpha_1+\alpha_2)}\Ltilde_{d-1}\left[\Delta\theta(\alpha_1-\alpha_2)\cos{\theta\over2}\right]e^{-2i(\alpha_1-\alpha_2)\sin{\theta\over2}}\ .}}
For $\tauhat\roughly<\Delta \tau$ this integral is easiest to perform by changing to the basis $\alpha_\pm=\alpha_1\pm\alpha_2$
where it is approximated by two independent one-dimensional Fourier transforms, resulting in an operator of size
\eqn\opantip{\langle0|\calO(b)|\psi_f\rangle\approx{(2\pi)^2 (\Delta\theta)^{d-1} \Nhat^2 \over 2^{2h_+-1} iC}(\omega R)^{4h_+-2}  
L\left({\tauhat\over\Delta\tau}\right)e^{-i\omega R\tauhat} {L_{d-1}\left({2\tan{\theta\over2}\over\Delta\theta}\right)\over \Delta\theta \cos{\theta\over 2}} }
where $L_{d-1}$ is the 1-dimensional inverse Fourier transform of $\Ltilde_{d-1}$. 
This result grows with $R$ at the antipodal point.  It falls as a Schwartz function in the immediate neighborhood of the antipode.
In deriving \opantip, in addition to $\tauhat\roughly<\Delta \tau$, we have assumed $1\gg \Delta \theta\gg\Delta \tau$.

\newsec{Scattering of boundary-compact wavepackets}

Now that we have found basic features of the wavepackets, we would like to see how closely boundary correlators, integrated against corresponding sources, can approximate the flat space S-matrix.  
While one could set up the problem of investigating scattering by deriving expressions like \psiscatt, given two ``in'' and two ``out'' wavepackets, we find that a useful and more intuitively clear way to proceed is to construct the ``in'' wavepackets, scatter them, and directly investigate properties of the resulting states.  Of course, it is then relatively simple to take such data and define an amplitude of the form \psiscatt, by integrating against the ``out'' wavepackets.  A particularly useful diagnostic for the  states is to look at the two-point function
%
\eqn\twooptwofield{\langle0|\calO(b_3)\calO(b_4)|\psi_{f_1}\psi_{f_2}\rangle\ .}
%
%
which ``register'' properties of the bulk state, in the boundary theory.
This can then be convolved with corresponding sources to find expressions of the form \psiscatt.

There are two contributions to \twooptwofield, the first being the direct contribution, where no scattering occurs, corresponding to the ``1'' in the S-matrix \Sdef, and the second is the scattered contribution. The direct contribution
\eqn\directcontrib{\langle0|\calO(b_3)\calO(b_4)|\psi_{f_1}\psi_{f_2}\rangle_d=\langle0|\calO(b_3)|\psi_{f_1}\rangle\langle0|\calO(b_4)|\psi_{f_2}\rangle}
is straightforward to examine using and \oppower\ and \opantip.
For example, outside of the near-forward region, we have power law falloff given by \oppower, 
\eqn\directcontribaway{\langle0|\calO(b_3)|\psi_{f_1}\rangle\langle0|\calO(b_4)|\psi_{f_2}\rangle\approx 
{\Ltilde(\omega\Delta t)^2[4\pi \nu  \Delta \tau (\Delta\theta)^{d-1} \Nhat{\hat L} \Ltilde_{d-1}(0)]^2\over(|\cos\theta_{13}-\sin\tauhat_3|)^{2h_+}(|\cos\theta_{24}-\sin\tauhat_4|)^{2h_+}}\ .}
Comparing this to the direct contribution for regular wavepackets in flat space, we find a marked difference. At angles away from the antipodal points to the sources, the direct contribution for regular wavepackets would fall as a Schwartz function, while for boundary-compact wavepackets we have found a power-law falloff.  One can likewise infer the behavior where one or both of the boundary points approaches an antipodal point, using \opantip.

We now turn to the scattered contribution to \twooptwofield, 
\eqn\scatteredcontrib{\langle0|\calO(b_3)\calO(b_4)|\psi_{f_1}\psi_{f_2}\rangle_s=i\gtilde^2\int d^Dx_1d^Dx_2G_{B\partial}(b_3,x_1)\psi_{f_1}(x_1)G_B(x_1,x_2)G_{B\partial}(b_4,x_2)\psi_{f_2}(x_2)\ }
where $\gtilde$ is the cubic coupling constant.
%

If we take $\tau_1\sim-\pi/2$ and $\tau_2\sim-\pi/2$, then $\psi_{f_1}(x_1)$ and $\psi_{f_2}(x_2)$ localize $x_1$ and $x_2$ in the approximately flat region \flatlim,\foot{Related discussion can be found in appendix B.} much as was described in the plane-wave limit in \GGP. Furthermore, we may choose to work in the center of mass frame by taking $\ehat_{10}=-\zhat$ and $\ehat_{20}=\zhat$. We write the bulk-boundary propagators as in \gbdelr, yielding
\eqn\scatcontribpsi{\eqalign{\langle0|\calO(b_3)\calO(b_4)|\psi_{f_1}\psi_{f_2}\rangle_s&\cr
\approx{i\gtilde^2\Nhat^2\over R^{d-1}}\int&d\alpha_1d\alpha_2d^Dx_1d^Dx_2(\alpha_1\alpha_2)^{2h_+-1}e^{i\alpha_1(\tau_1-\tauhat_3-\vecx_1\cdot\ehat_3)}e^{i\alpha_2(\tau_2-\tauhat_4-\vecx_2\cdot\ehat_4)}\cr&G_B(x_1,x_2)\psi_{f_1}(x_1)\psi_{f_2}(x_2)\ .}}
Using the form of $\psi_f$ found in \flatpack\ and approximating the bulk propagator by the flat space expression,
%
\eqn\Gflat{G_B(x_1,x_2)\approx \int{d^Dp\over(2\pi)^D}{e^{ip\cdot(x_1-x_2)}\over p^2-i\epsilon}\ ,}
we find
\eqn\sccontribcenter{\eqalign{&\langle0|\calO(b_3)\calO(b_4)|\psi_{f_1}\psi_{f_2}\rangle_s
\approx{i\gtilde^2(\Delta t)^2\Nhat^4 R^{4\nu}}\int d^d\veck_1d^d\veck_2dk_3dk_4(k_1k_2)^{2h_+-d}(k_3k_4)^{2h_+-1}\cr
&\times(2\pi)^D\delta^D(k_1+k_2-k_3-k_4){e^{-iR(k_3{\tauhat_3}+k_4{\tauhat}_4)}\over(k_1-k_3)^2}
\Ltilde(\Delta t(k_1-\omega))L({\theta_1\over\Delta\theta})\Ltilde(\Delta t(k_2-\omega))L({\theta_2\over\Delta\theta})}\ ,}
where we have identified
\eqn\kthreefour{\veck_3={\alpha_1\over R}\ehat_3\ \ \ \ \ \veck_4={\alpha_2\over R}\ehat_4}
and defined $\cos\theta_i=\ehat_i\cdot\ehat_{i0}$. 

We would like to understand how this differs from the flat-space scattering amplitude for regular wavepackets. Before considering wavepackets, let us first consider hypothetically replacing the incoming states by definite momentum states,  
\eqn\Ltodelta{\Ltilde(\Delta t(k_i-\omega))L({\theta_i\over\Delta\theta})\rightarrow {(\Delta\theta)^{d-1} \Ltilde_{d-1}(0)\over \Delta t} \delta(k_i-\omega)\delta^{d-1}(\ehat_i -\ehat_{i0})}
and recall that we have chosen to work in the center of mass frame by placing our sources such that $\ehat_{1,0}\cdot\ehat_{2,0}=-1$. This gives 
%
\eqn\planewavescat{\eqalign{\langle&0|\calO(b_3)\calO(b_4)|\psi_{f_1}\psi_{f_2}\rangle_s\quad\cr
&\approx i (g^2\omega^{d-5}) \Nhat^4 (\omega R)^{4\nu}   {(2\pi)^D\delta^{d-1}(\ehat_3+\ehat_4) 
 \over2\sin^2{\theta_{13}\over2}}
  e^{-i\omega R(\tauhat_3+\tauhat_4)}    [(\Delta\theta)^{d-1} \Ltilde_{d-1}(0)]^2}}
which is the usual flat space scattering result for plane waves, up to normalization factors due to the AdS construction. 

However, our incoming states can not have a definite momentum. Specifically, while the boundary-compact wavepackets we have constructed are approximately localized to the flat region \flatlim, they have low-energy tails of order $\Ltilde(\omega\Delta t)$, which can lead to significant effects. To see these effects, consider wavepackets at definite angles, {\it i.e.} with $\Delta \theta\rightarrow 0$, but with a general energy profile: 
\eqn\LtoReg{\Ltilde(\Delta t(k_i-\omega))L({\theta_i\over\Delta\theta})\rightarrow f_i(\Delta t(k_i-\omega))\delta^{d-1}(\ehat_i-\ehat_{i0})(\Delta\theta)^{d-1} \Ltilde_{d-1}(0)\ .}
Making this replacement, we have 
\eqn\defanglscat{\eqalign{\langle0|\calO(b_3)\calO(b_4)|\psi_{f_1}\psi_{f_2}\rangle_s\quad&
{\buildrel\sim\over\rightarrow}\quad{i\gtilde^2(\Delta t)^2\Nhat^4R^{4\nu}(2\pi)^D}\delta^{d-2}(k_{4\perp})[(\Delta\theta)^{d-1}\Ltilde_{d-1}(0)]^2\ \times\cr\int dk_3 dk_4 &e^{-iR(k_3{\tauhat}_3 + k_4 {\tauhat}_4)} (k_1k_2k_3k_4)^{2h_+-3/2}\cr
& f_1[\Delta t(k_1-\omega)]  f_2[\Delta t(k_2-\omega)] {\delta(k_3 \sin\theta_3 -k_4\sin\theta_4)\over2\sin(\theta_3/2) \sin(\theta_4/2)}\ ,}}
where $\theta_3$ is the angle between $k_3$ and $k_1$, $\theta_4$ is the angle between $k_4$ and $k_2$, 
$k_1$ and $k_2$ satisfy 
\eqn\momconsol{k_1=k_3 \cos^2{\theta_3\over 2} + k_4 \sin^2{\theta_4\over 2}\quad,\quad k_2=k_3 \sin^2{\theta_3\over 2} + k_4 \cos^2{\theta_4\over 2}\ ,}
and where $k_{4\perp}$ is the component of $k_4$ perpendicular to the scattering plane defined by the other three momenta.

In the case of regular wavepackets, where $f_i(\Delta t(k_i-\omega))$ are compactly supported on the momentum interval $[\omega-{1\over\Delta t},\omega+{1\over\Delta t}]$ and $\omega\Delta t>1$, while there are non-zero contributions away from $\theta_3=\theta_4$, the largest deviation is
\eqn\regulardev{|\theta_3-\theta_4|_{max}\simeq {\sin\theta_3 + \sin\theta_4\over \omega\Delta t}\ .}
%
For $|\theta_3-\theta_4|>|\theta_3-\theta_4|_{max}$, the scattered contribution for regular wavepackets vanishes identically. 

However, we are unable to construct regular wavepackets from boundary-compact sources, and instead have $f_i=\Ltilde[\Delta t(k_i-\omega)]$. Thus
\eqn\irregscat{\eqalign{\langle0|\calO(b_3)&\calO(b_4)|\psi_{f_1}\psi_{f_2}\rangle_s
{\buildrel\sim\over\rightarrow}{i\gtilde^2\Nhat^4(\Delta t)^2[(\Delta\theta)^{d-1}\Ltilde_{d-1}(0)]^2\left(\cos{\theta_3-\theta_4\over2}\right)^{4h_+-3}R^{4\nu}\over2\sqrt{\sin\theta_3\sin\theta_4}\sin{\theta_3\over2}\sin{\theta_4\over2}}(2\pi)^D\times \cr
\delta^{d-2}(k_{4\perp})
&\int d\kappa\kappa^{8h_+-6}\Ltilde\left[\Delta t(\kappa\zeta_1-\omega )\right]\Ltilde\left[\Delta t(\kappa\zeta_2-\omega )\right]
e^{-iR\kappa\left(\tauhat_3\sqrt{\sin\theta_4\over\sin\theta_3}+\tauhat_4\sqrt{\sin\theta_3\over\sin\theta_4}\right)}\ ,
}}
where
\eqn\zetadef{\zeta_1=\sqrt{\sin\theta_4\over\sin\theta_3}\cos^2{\theta_3\over2}+\sqrt{\sin\theta_3\over\sin\theta_4}\sin^2{\theta_4\over2}\quad,\quad\zeta_2=\sqrt{\sin\theta_4\over\sin\theta_3}\sin^2{\theta_3\over2}+\sqrt{\sin\theta_3\over\sin\theta_4}\cos^2{\theta_4\over2}\ .}

First consider $\theta_3\sim\theta_4$, $\zeta_i\approx1$, so for $\tauhat_{3,4}\ll\Delta \tau$ both $\Ltilde$ are equally peaked, fixing $\kappa\approx\omega $. Then the dominant contribution comes from the center of both wavepackets, as in the case of regular wavepackets.  Specifically, for $\theta_3=\theta_4$ this gives the expression
\eqn\smalltau{\langle0|\calO(b_3)\calO(b_4)|\psi_{f_1}\psi_{f_2}\rangle_s
\approx {i\gtilde^2\Nhat^4[(\Delta\theta)^{d-1}\Ltilde_{d-1}(0)]^2\over (\Delta t)^{8h_+-7}} {R^{4\nu}(2\pi)^D
\delta^{d-2}(k_{4\perp})\over 2 \sin\theta_3 \sin^2{\theta_3\over 2} }e^{-i\omega R(\tauhat_3+\tauhat_4)} {\hat L}_2}
where $\hat L_2$ is a number determined by a Fourier transform of $\Ltilde^2$.

But, there are also significant scattered contributions  for $\theta_3\neq\theta_4$.  For example, 
 if $\theta_3\ll\theta_4$, then $\zeta_1\gg\zeta_2$, and so the first $\Ltilde$ is more sharply peaked, fixing $\kappa\approx\omega /\zeta_1$, and so
%
%
\eqn\haze{\eqalign{\langle0|\calO(b_3)&\calO(b_4)|\psi_{f_1}\psi_{f_2}\rangle_s
{\buildrel\sim\over\rightarrow}{i\gtilde^2\Nhat^4(\Delta t)[(\Delta\theta)^{d-1}\Ltilde_{d-1}(0)]^2\check LR^{4\nu}\left(\cos{\theta_3-\theta_4\over2}\right)^{4h_+-3}\over2\sqrt{\sin\theta_3\sin\theta_4}\sin{\theta_3\over2}\sin{\theta_4\over2}}\times \cr
&\Ltilde(\omega\Delta t){\omega^{8h_+-6}\over \zeta_1^{8h_+-5}} (2\pi)^D\delta^{d-2}(k_{4\perp})
e^{-i{R\omega\over\zeta_1}\left(\tauhat_3\sqrt{\sin\theta_4\over\sin\theta_3}+\tauhat_4\sqrt{\sin\theta_3\over\sin\theta_4}\right)}\ ,
}}
where $\check L$ is a numerical coefficient arising from another Fourier transform of $\Ltilde$.
We find a similar result for $\theta_3\gg\theta_4$, with the roles of $\zeta_1$ and $\zeta_2$ reversed. This has support of order $\Ltilde(\omega\Delta t)$ for $|\theta_3-\theta_4|$ of order 1 and falls as a power of $\cos{\theta_3-\theta_4\over2}$ for $|\theta_3-\theta_4|\rightarrow\pi$, contrasting sharply with the case of regular wavepackets, where the amplitude would be identically zero in this regime. 

To physically understand how the non-compact momentum tails lead to this order $\Ltilde(\omega\Delta t)$ haze of scattered particles at all angles, consider the two cases $k_1>k_2$ and $k_1<k_2$. For $k_1>k_2$, $\psi_1$ will scatter less than the $k_1=k_2$ case, while $\psi_2$ will scatter more, causing $\theta_{13}$ and $\theta_{24}$ to be smaller and larger than expected, respectively. 
For $k_1<k_2$, roles reverse.

Both this haze of scattered particles, and competition between the direct and scattered contributions, raise questions about our ability to sharply resolve bulk physics, to which we turn next.


\newsec{The question of resolution}

If one sought to give a complete non-perturbative definition of string theory via a dual CFT, a starting point for such a definition is the set of correlators for the theory, which one might imagine computing, {\it e.g.}, on a sufficiently large computer.  The full correlators include contributions from what, in the bulk viewpoint, are the direct and scattered contributions:
\eqn\boundamp{
\eqalign{\int db_1 db_2 f_1(b_1) f_2(b_2) \langle0|\calO(b_3)&\calO(b_4)\calO(b_1)\calO(b_2)|0\rangle
\cr&= \langle0|\calO(b_3)\calO(b_4)|\psi_{f_1}\psi_{f_2}\rangle_d+\langle0|\calO(b_3)\calO(b_4)|\psi_{f_1}\psi_{f_2}\rangle_s\ .}}
A question is how accurate of an approximation to the elements of the flat-space S-matrix can be derived from such correlators.  In particular, to approximate flat
space scattering of Schwartz or regular wavepackets, we should recover a corresponding expression for the S-matrix of the form given in section 2. Instead, our boundary construction produces amplitudes given by \boundamp, which differ from that form in the direct and scattered contributions. Let us examine these results more closely to determine in what regimes  the boundary amplitude closely approximates \psiscatt, for a basis of Schwartz or regular wavepackets. 

%

First, consider just the scattered contribution, as described at the end of the preceding section.  There we find an additional haze of scattered particles due to the low-momentum tails. 
%
In particular, the ratio of the amplitude for the haze at $\theta_3\neq\theta_4$, \haze, to the contribution at $\theta_3=\theta_4$, \smalltau, is of size
\eqn\hazeiss{(\omega \Delta t)^{8h_+-6} \Ltilde(\omega \Delta t)\ .}
Thus wavepackets with $\omega\Delta t\sim \calO(1)$
receive significant corrections to their scattering, compared to the usual wavepackets.   In this range, amplitudes for the boundary-compact wavepackets appear not to sharply reproduce the flat S-matrix elements, \psiscatt, of
Schwartz or regular wavepackets.  (The limit of \GGP\ corresponds to $\omega\Delta t\rightarrow \infty$, where \hazeiss\ vanishes.)

%
%
To recover the full S-matrix in the limit $R\rightarrow\infty$, we must have a sufficiently complete set of wavepackets. However, such a set must apparently include wavepackets with $\omega\Delta t \sim {\cal O}(1)$, in which case $\Ltilde(\omega\Delta t)$ is not small and the tail effects compete with the desired scattering amplitude, impeding the attempt to recover the flat space S-matrix. Put differently, if $\Delta t$ represents the desired size of the lab, we have encountered an apparent limitation in reproducing the S-matrix when considering wavepackets with comparable wavelengths.

If one notes that in a derivation of correlators within a CFT, one computes the sum of the direct and scattered contribution, 
\boundamp, we see potential for a more serious issue.  Specifically, due to the tails, the direct contribution is not sharply restricted to the forward direction.  For example, if we consider taking the boundary times to lie in the small interval of size $\Delta \tau$, which was in effect used to extract the plane-wave limit of the S-matrix elements in \GGP\ (where the direct contribution was not included), we find that the direct contribution to the correlator gives an angular distribution given by a product \directcontrib\ of two terms of the form \opantip.  This is to be compared to the scattered contribution, which in the same regime is given by \smalltau, \haze.  Comparing the magnitude of the two angular distributions gives
\eqn\magnitudrat{{\langle0|\calO(b_3)\calO(b_4)|\psi_{f_1}\psi_{f_2}\rangle_s\over \langle0|\calO(b_3)\calO(b_4)|\psi_{f_1}\psi_{f_2}\rangle_d}\sim  {(g^2\omega^{d-5}) (\omega R)^{4-2d}\over (\omega \Delta t)^{(8h_+-7)}}\ .}
Thus, in the large-$R$ limit the strength of the scattered signal is suppressed relative to the tail of the direct signal, raising the question of how one can isolate the scattered signal and extract the relevant elements of the S-matrix.

Note that in the boundary theory, this suppression can be interpreted as being largely due to a relative $1/N^2$ between the two contributions.  This suggests a possibility of finding a way to separate the two contributions that is intrinsic to the CFT, through the systematics of the $1/N^2$ expansion (or perhaps via the singularities discussed in \GGP).  Here, though, note that one seemingly doesn't necessarily have the advantages available in {\it e.g.} Rutherford scattering, where the subleading contribution in the weak-coupling expansion can be isolated from the direct contribution to scattering by considering detectors at finite angle, where the direct contribution is negligible.

The effects we have discussed arise from the tails of our boundary-compact wavepackets, and one might take the viewpoint that the tails are an inessential complication.  After all, one can take the limit 
$\omega\Delta t\rightarrow\infty$ to recover plane waves, and indeed \GGP\ proposed a prescription by which bulk S-matrix elements might in-principle be recovered in this limit.  One could then imagine superposing these to make an arbitrary wavepacket.

A more careful phrasing of the question is the following.  Our boundary-compact construction allows us to construct a certain space of wavepackets.  If we are to recover the complete S-matrix, an important question is whether these wavepackets are dense in the desired Hilbert space of single-particle states.

While we do not have a complete answer to this question we can see that 
that matters may not be so simple turning the question around.  Suppose that we have a wavepacket in the bulk of AdS that is a Schwartz function;\foot{Note that such a function will not necessarily remain Schwartz in AdS, but we avoid this matter by focusing on sufficiently short time scales.} specifically, suppose that it is taken to have exponential falloff, of the form
\eqn\bulksch{\psi_f(x) \propto \exp\left\{ -\left({r\over r_0}\right)^a \right\} }
for some width $r_0$.  In that case, for finite $R$, the size of this wavepacket at the radius $r\sim R$ where the AdS geometry begins to be relevant, is exponentially small in $R$.  This in turn means that when we take the scaling limit \boundcorr\ to find the corresponding behavior of the boundary operator, we have
\eqn\schwbdy{\langle 0|\calO |\psi_f\rangle \sim e^{-(R/r_0)^a}\ .}
Recall that $R^4 \propto N l_p^4$, where $N$ is the number of branes, and $l_p$ the Planck length.  From this, we see that the boundary operator is nonperturbatively small in $1/N$, suggesting that we need non-perturbative control over the theory to exhibit localized wavepackets and study their tree-level scattering, and that in any case such wavepackets don't trivially follow from superposing those that we have considered.  Indeed, the form \schwbdy\ is suggestive of the size of matrix elements that one might expect if the wavepacket corresponds to excitation of the $N^2$ matrix degrees of freedom in a non-trivial way.  Note moreover that transversely focussed wavepackets, with $\omega\Delta x_\perp\sim {\cal O}(1)$, must have $\Delta\theta\sim{\cal O}(1)$ and thus be spread over a significant fraction of the boundary $S^3$, also possibly indicating a role for matrix degrees of freedom. Of course, this is also suggested by the observation that the Bekenstein-Hawking entropy of an AdS region of size $R$ is $S_{BH}\sim N^2$. If this is the situation, we see no guarantee that the corresponding excitations  will produce the correct S-matrix elements of a local bulk theory, for such states.  The question is even more sharply stated if one insists on being able to represent compact-support wavepackets in the bulk.  Via the prescription \boundcorr, we see that these would have to correspond to  boundary states with {\it vanishing} value for $\langle0|{\cal O}(b)|\psi_f\rangle$.

\newsec{Discussion}

We have found that the wavepackets considered in \GGP, which yield formulas for S-matrix elements in the plane wave limit, do not accurately approximate all scattering amplitudes of wavepackets of rapid decrease (Schwartz functions) or compact support.  This behavior arises due to certain long-distance power-law tails, which correspond in momentum space to low-energy tails.  The lack of accurate approximation becomes particularly acute for wavepackets whose width and wavelength are comparable.  This suggests that probes of bulk physics through such wavepackets could lack a certain resolution.

One might ask whether this is simply a problem requiring introduction of more general wavepackets.  Let us review the arguments that matters may not be so simple.  First, data specified at timelike infinity does not lead to wavepackets that asymptotically separate -- instead they scatter infinitely many times.  Second, data on the spacelike boundary of AdS in general leads to divergences and delocalization, since such sources are convolved with correlators associated with non-normalizable bulk behavior.  The only obvious way to avoid such behavior is to restrict ones boundary sources to have compact support, which is precisely what we have done.

Moreover, we have briefly investigated more generally what boundary one-point functions would correspond to bulk wavepackets of rapid decrease, or those of compact support.  We find that the former seem to have corresponding boundary correlators indicative of non-trivial excitation of the $N^2$ matrix degrees of freedom, and the latter have vanishing value for the one-point function corresponding to the field in question.  These statements suggest to us that full recovery of scattering amplitudes for such wavepackets might require additional non-trivial behavior of the $N^2$ matrix degrees of freedom, and if so, we see no principle that guarantees that such behavior arises.  Thus, in short, our arguments raise the question of how extraction of 
of the corresponding elements of the scattering matrix is possible, from the correlators of the boundary theory, in the absence of rather non-trivial effects.

If the boundary theory did not fully specify a unitary S-matrix for the bulk theory, then this could have important consequences.  For example, claims that the AdS/CFT correspondence solves the black hole information paradox appear to rest on this sort of complete unitary description of the bulk theory.\foot{Indeed, we note that one expects typical Hawking quanta to be emitted in wavepackets whose size is comparable to their wavelength, $\omega\Delta t\sim 1$, both being comparable to the black hole radius, which is the regime where boundary-compact data least accurately reproduces usual flat S-matrix elements.}   If this were the case, the AdS/CFT correspondence might only permit study of certain gross features of black hole physics.
 
More generally, there are now apparently two non-trivial tests for whether a given CFT reproduces a bulk S-matrix and unitary bulk evolution with familiar features such as approximate locality, {\it etc.}  The first test arises from \GGP:  it appears that the boundary CFT has to have certain characteristic singularity structure in its correlators, to produce the plane-wave limit of the S-matrix.  Secondly, one must have a way of specifying boundary data that allows one to construct the appropriate complete set of asymptotic scattering states of the bulk theory.  These thus both appear to be tests that can be used in diagnosing whether a particular CFT accurately reproduces a bulk theory that has expected properties, such as locality on scales large as compared to the string and Planck lengths, in situations where scattering is not ultraplanckian\refs{\GiSr,\GiPo}.  

If no CFT did fully reproduce an appropriate bulk S-matrix, such as that of string theory, one might ask how the AdS/CFT correspondence could have had as much success as it has had in matching certain bulk features to boundary features.  We do not have a complete answer, but two common themes throughout physics are the power of {\it symmetry} and {\it universality}.  Perhaps successful matches between bulk and boundary physics might be analogous to other situations where  theories that are inequivalent on a fine-grained level nonetheless match in many features of their coarse-grained behavior, as is {\it e.g.} commonly seen in effective field theory.  One might contemplate investigating how robust the boundary predictions of AdS/CFT, typically obtained in the form of boundary correlators, would be to adjustments of the fine details of the bulk theory, {\it e.g.} at scales $\roughly< R$; the required singular behavior of \GGP\ does suggests a certain such sensitivity, but it may be that accessible features of the boundary theory are  insensitive to certain perturbations of the bulk physics.  In any case, we regard these questions regarding possible limitations on resolution of the AdS/CFT hologram as deserving more complete answers.

\bigskip\bigskip\centerline{{\bf Acknowledgments}}\nobreak

We have  benefited from conversations with L. Cornalba, B. Gripaios, J. Polchinski, R. Stora, D. Trancanelli, and E. Witten.  We also acknowledge  many discussions with J. Penedones, in the course of preparing \refs{\GGP}, and D. Marolf both for discussions, and for comments on a draft of this work.  This work was supported in part by the U.S. Department of Energy under Contract DE-FG02-91ER40618, by grant RFPI-06-18 from the Foundational Questions Institute (fqxi.org), and by a Marie Curie Early Stage Research Training Fellowship of the European Community's Sixth Framework Programme under contract number MEST-CT-2005-020238-EUROTHEPHY.  We gratefully acknowledge the hospitality of the CERN theory group, where part of this work was carried out.

\appendix{A}{AdS propagators}

In this appendix we collect some facts about propagators. In particular, it is important to have a careful definition of the $i\epsilon$ prescription. Following \refs{\BuLu}, we write the propagator as a sum over normalizable solutions to the Klein-Gordon equation with Feynman boundary conditions. We make use of the homogeneity of AdS to choose $x'=0$:
\eqn\GBSum{G_B(x,0)={i e^{-i2h_+|\tau|}(\cos\rho)^{2h_+}\over 2R^{d-1}\pi^{d/2}}\sum_{n=0}^\infty{\Gamma(2h_++n) \over \Gamma(n+\nu+1)}e^{-2ni|\tau|}
                             P_n^{(d/2-1,\nu)}(\cos2\rho)\ .}
Written in this form, the propagator manifestly exhibits periodicity in $\tau$ with period $2\pi$, up to an overall phase $e^{-4\pi ih_+}$, as inherited from the embedding description of AdS. Furthermore, as required by Feynman boundary conditions, the propagator is manifestly symmetric under time-reversal and purely positive frequency in the future. If we deform $|\tau-\tau'| \mapsto |\tau-\tau'|-i\epsilon$ this sum converges to\refs{\Henrici}
\eqn\GB{G_B(x,x')={i\Cprop \over R^{d-1}(1+\sigma_\epsilon/R^2)^{2h_+}}F\left(h_+,h_++{1\over2};\nu+1;{1 \over (1+\sigma_\epsilon/R^2)^2}\right)}
where 
\eqn\Geodetic{\sigma_\epsilon(x,x')=R^2\left(-1 + {\cos(|\tau-\tau'|-i\epsilon)-\sin\rho\sin\rho' \ehat\cdot\ehat' \over \cos\rho\cos\rho'}\right)}
is the geodetic distance, defined with respect to the embedding metric, and we have defined the constant 
\eqn\cBdef{\Cprop={\Gamma(2h_+) \over 2^{2h_++1} \pi^{d/2} \Gamma(\nu+1)}\ .}
%
This is the form of the propagator found in \refs{\DulB,\Dohse}, although the $i\epsilon$ prescription has been made more transparent. Using the quadratic transformation for hypergeometric functions, (15.3.20) in \refs{\AbSt}, we can further simplify this expression to
\eqn\GB{G_B(x,x')={i\Cprop \over R^{d-1}(\sigma_\epsilon/R^2)^{2h_+}}F\left(2h_+,{2\nu+1\over 2};2\nu+1;{-2R^2 \over \sigma_\epsilon}\right)\ .}
This particularly simple form of the propagator with $i\epsilon$ prescription is to our knowledge not present elsewhere in the literature. 

There are a number of further checks we can perform to ensure we have, indeed, found the correct form of the propagator. Some of these checks, including ensuring the propagator has the correct equal-time behavior, have been performed in \refs{\Dohse}, although some properties were left uncofirmed. We can consider the propagator at scales much smaller than $R$, in which case we find
\eqn\GBShortDist{G_B(x,x')\approx {i\Gamma({d-1\over2}) \over 4\pi^{(d+1)/2}}{1\over\left[-(t-t')^2+(\vecx-\vecx')^2+i\epsilon\right]^{d-1\over2}}}
which is the expected flat space behavior. Taking one point to the boundary and rescaling as in \gbbdef, we find the bulk-boundary propagator
\eqn\GBd{G_{B\partial}(b,x')=\lim_{\rho\rightarrow{\pi\over2}} 2\nu R^{d-1\over2}{G_B(x,x') \over (\cos\rho)^{2h_+}}={2i\nu\Cprop\over R^{d-1\over2}}\left({\cos\rho' \over \cos(|\tau-\tau'|-i\epsilon)-\sin\rho'\ehat\cdot\ehat'}\right)^{2h_+}\ .}
Taking the other point to the boundary and again rescaling as in \gbbdef, we find the boundary CFT propagator
\eqn\Gd{G_{\partial}(b,b')=\lim_{\rho'\rightarrow{\pi\over2}} 2\nu R^{d-1\over2}{G_{B\partial}(b,x') \over (\cos\rho')^{2h_+}}={i(2\nu)^2\Cprop\over \left[\cos(|\tau-\tau'|-i\epsilon)-\ehat\cdot\ehat'\right]^{2h_+}}\ .}
%

%
%
%
%

%
%
%
%

\appendix{B}{Localization in a flat region}

It is useful to understand the boundary image of a state that is an  approximate plane wave localized to one approximately flat region, given on the spatial slice $t=0$ by
\eqn\approxplanewave{\Psi(\veck,\vecx_0)=L\left({|\vecx-\vecx_0|\over\Delta x}\right)e^{ik\cdot(x-x_0)}}
where $\Delta x\ll R$, localizing $\Psi$ to a single region near $\vecx_0$ of size much less than $R$. Such a state propagates to the boundary according to
\eqn\opplanewave{\eqalign{\langle0|\calO(b)|\Psi\rangle\approx&{\Nhat \over R^{d-1\over2}}\int d\alpha{d^d\vecx'}\left.\alpha^{2h_+-1}L\left({|\vecx'-\vecx_0|\over\Delta x}\right)e^{-ikt'+i\veck\cdot(\vecx'-\vecx_0)}(-i\lrd_{t'})e^{i\alpha(\cos|\tau-\tau'|-{\vecx'\over R}\cdot\ehat)}\right|_{\tau'=0}\cr
\approx&{i\Nhat ({\Delta x})^d\over R^{d-1\over2}}\int d\alpha\alpha^{2h_+-1}(k+{\alpha\over R}\sin\tau)\Ltilde_d\left(\Delta x\left|\veck-{\alpha\over R}\ehat\right|\right)e^{i\alpha(\cos\tau-{\vecx_0\over R}\cdot\ehat)}\ .}}

If we take $|\vecx_0|\ll R$, restricting ourselves to contributions from the flat region \flatlim, \opplanewave\ is similar to 
expressions considered in section 4; in particular, for $\Delta x k\gg1$, it is well-localized in momentum, $\alpha\ehat \simeq {\vec k}R$.  Note that if $\vecx_0$ is moved away from the center of the AdS region in question, this can produce an amplitude at the boundary at an angle that is not the same as the scattering angle.  
Specifically, if instead of scattering near $|\vecx_0|=0$, the tail of one of the wavepackets scatters off the center of another wavepacket at some non-zero radius $|\vecx_0|$ at an angle $\theta_0$, the scattered amplitude will arrive at the boundary at an angle
\eqn\offcenterangle{\theta=\sin^{-1}\left({\sin{|\vecx_0|\over R}\tan\theta_0 \over \sqrt{({|\vecx_0|\over R})^2\sec^2{|\vecx_0|\over R}+\sin^2{|\vecx_0|\over R}\tan^2\theta_0}}\right)\ .}
For $|\vecx_0|\sim R$, $\theta-\theta_0$ can be of order 1. While this might have seemed to lead to additional loss of resolution,  in this region the wavepacket tail is of order ${\psi_f(0)\Ltilde(\omega\Delta t)\over(\omega R)^{2h_+}}$. Thus these contributions are supressed by additional powers of $R$.

\listrefs

\end